\documentstyle{article}
\input math_macros
\font\twelvebf=cmbx12
\font\twelverm=cmr12
\font\twelveit=cmti12

\font\tenrm=cmr10

\font\elevenbf=cmbx10 scaled\magstep 1
\font\elevenrm=cmr10 scaled\magstep 1
\font\elevenit=cmti10 scaled\magstep 1

\renewenvironment{thebibliography}[1]
 { \elevenrm
   \begin{list}{\arabic{enumi}.}
    {\usecounter{enumi} \setlength{\parsep}{0pt}
     \setlength{\itemsep}{3pt} \settowidth{\labelwidth}{#1.}
     \sloppy
    }}{\end{list}}
\begin{document}
\begin{center}{{\twelvebf EFFECTIVE FIELD THEORY\\
               \vglue 14pt
               FOR DIFFRACTIVE QCD PROCESSES\\}
\vglue 1cm
{\twelverm SOO-JONG REY\\}
\baselineskip=18pt
{\twelveit Physics Department \& Center for Theoertical Physics\\}
\baselineskip=12pt
{\twelveit Seoul National University, Seoul 151-742 KOREA\\}
\vglue 0.6cm
{\tenrm ABSTRACT}}
\end{center}
{\rightskip=3pc
 \leftskip=3pc
 \elevenrm
 \baselineskip=12pt
 \noindent
An effective field theory describing the QCD diffraction scattering
is constructed. The constituent quarks interacting
with multiperipheral gluons and Goldstone bosons are
described by impact-parameter two-dimensional sigma models
for the color and the flavor degrees of freedom.
The elastic scattering amplitude is then shown to factorize into
products of
hard and soft pomeron contributions.
\vglue 0.6cm}
\begin{center}
{{\twelveit Talk at 5th Blois Workshop on Elastic and Diffractive
Scattering\\
\vglue 12pt
\twelveit June 8 - 12, 1993, Providence RI USA\\}
}
\end{center}
\twelverm
\baselineskip=23pt

QCD has proven to be the right description of the strong interaction
at short distances, yet the evidence is not so much for the
high-energy diffractive processes. Current data appears to fit with
vectorlike `pomeron' exchanges with $C=+1$ and isosinglet
quantum numbers \cite {lipatov}, as a result of complicated
low-energy nonperturbative QCD dynamics.
As such, by varying the momentum transfer $t$
near the confinement and chiral symmetry breaking
$(\Lambda_{\rm qcd} \approx 0.2 GeV,
\Lambda_{\rm csb} \approx 1.2 GeV)$, constitutents of
the pomeron is suspected to change from a `soft' to a `hard' ones.
In this talk, I report my own works \cite {rey} on constructng an
effective field theory (EFT) describing the
diffractive interactions of quarks and gluons starting from the QCD.

The relevant degrees of freedom are the constituent light quarks
$Q \equiv (u, d, s)$
surrounded by gluons and Goldstone bosons of the chiral symmetry
breaking as studied by Georgi, Manohar and Weinberg
\cite {georgimanohar}, \cite {weinberg}. This viewpoint is
motivated by the highly successful `additive quark rule'
in the non-annihilation channel, point-like Dirac particle behavior
of the constituent quarks \cite {weinberg}, and the model's easy
account of Ellis-Jaffe spin and Gottfried isospin sum-rule
violations \cite {bjorken} \cite{eichten}.
I start from the QCD in which the Goldstone bosons are explicitly
taken into account.
After integrating out the quark and gluon high-mometum
components above the chiral symmetry breaking scale $\Lambda_{\rm csb}$,
the EFT of the constituent quark reads:
$$
\eqalign{
L_4 = & -{1 \over 4 g^2} tr G_{\mu \nu} G^{\mu \nu} +
\bar Q (i {\cal D} \hskip-0.23cm / - M_Q ) Q
+ {f^2 \over 4} Tr \nabla_\mu \Sigma^\dagger \nabla^\mu \Sigma \cr
& + L_1 Tr (\nabla_\mu \Sigma^\dagger \nabla^\mu \Sigma)^2
+ L_2 (Tr \nabla_\mu \Sigma^\dagger \nabla_\nu \Sigma)^2
+ L_3 Tr (\nabla^\mu \Sigma^\dagger \nabla_\nu \Sigma)^2
+ \cdots\cr
}
\eqno (1)
$$
Notations are as follows. $G_{\mu \nu}$ denotes the gluon field
strength;
$\Sigma \equiv \xi^2$, $ \xi \equiv \exp ( i \Pi / f)$,
$f \approx 93 MeV$, $\Pi$ defines the Goldstone boson fields
which transform as $\Sigma \rightarrow L \Sigma R^\dagger$ and
$\xi \rightarrow L \xi U^\dagger = U \xi R^\dagger$ under
$SU_L (2) \otimes SU_R (2)$ chiral transformations;
the covariant derivative $i{\cal D}_\mu \equiv i D_\mu^{(\rm color)}
+ V_\mu + \gamma_5 A_\mu$,
$V_\mu = {1 \over 2} (\xi^\dagger \partial
\xi + \xi \partial \xi^\dagger)$, $A_\mu \equiv { i \over 2}
(\xi^\dagger \partial \xi - \xi \partial \xi^\dagger)$ in which
the axial coupling is taken to unity \cite {weinberg}.
This theory has already shown a good success in
explaining the static properties of
the low-energy QCD such as hadron spectroscopy \cite {georgimanohar}.
In addition, fluctuation of the parton distributions is assumed
well described by the interactions of
constituent quarks with Goldstone bosons and gluons. Having
included both the Goldstone boson and gluon degrees of
freedom simultaneously, we expect the theory Eq.(1) provides
a reasonable starting point of investigating the nature of pomerons,
in particular, possible distinct experimental signatures
between the `soft' and the `hard'
pomerons if any. This is the issue of immediate interest
and importance to the SSC/LHC experiments \cite{bjorken}.

Consider a general $2 \rightarrow 2 + n$ scattering of quarks
at high energy. In the center of mass frame,
each constituent quarks inside colliding hadrons are boosted
to 4-velocities $n^\mu \equiv (1, \vec P/E)$
and $ \bar n^\mu \equiv (1, - \vec P / E)$ respectively.
Both 4-velocities are null; $n^2 = \bar n^2 = 0$,
$n \cdot \bar n = 2$. In the diffractive process, the constituent
quarks become eikonal, and the gluon and (longitudinal gradients of)
the Goldstone boson fields seen by them are Lorentz contracted.
It is possible to construct a  new EFT incorporating
these features manifestly.
Consider small fluctuations of the high-energy boosted
quark about the four-momentum $P^\mu = E n^\mu$.
The fluctuating quarks' four-momentum may be denoted as
$P_\mu = E n_\mu + k^\mu$,
in which $k^\mu, M_Q <\!\!< E$, $k \cdot n = 0$.
Cell-decomposing each small fluctuations
with new effective fermion fields labelled by 4-vector $n^\mu$,
$$
Q_\pm (x) \equiv {1 \pm \gamma_5 \over 2} Q(x) =
\sum_{(n^2 = 0)} e^{- i E n \cdot x } Q_{n \pm} (x)
= \sum_{(n^2 = 0)} \int {d^4 k \over (2 \pi)^4} \delta (n \cdot k)
e^{- i En \cdot x - i k \cdot x} \tilde Q_{n \pm} (x).
\eqno (2)
$$
The new EFT of high-energy constituent quarks
$Q_{n \pm}$ is then obtained by inserting Eq.(2) into
Eq.(1) and expanding in powers of $1/E$:
$$
L_{\rm Q} = \!\!\sum_{\{n\}, \pm} \!\!Q_{n\pm}^{\dagger} in\!\cdot\!
i {\cal D}_{n \pm} Q_{n \pm}
-{1 \over 2 E} Q^\dagger_{n\pm} {\cal D}^2_\pm Q_{n \pm}
-{i \over 2 E} Q^\dagger_{n\mp} (\sigma_{\mu \nu} + n_\mu \gamma_\nu)
G^{\mu \nu} Q_{n \pm}
+ (n^\mu \leftrightarrow \bar n^\mu) + \cdots
\eqno (3)
$$
Not only is the new EFT simplified, but also emerges extra symmetries
not present in the underlying QCD.
The leading order, kinetic term of the constituent quark is
independent of its helicity.
Therefore it shows (1) scaling symmetry, $ E \rightarrow
E + \delta E$, and (2) helicity-flavor $SU(2 N_f)$ symmetry for each
light-cone direction $n^\mu$. These symmetries are broken by various
other interactions: the Goldstone boson interactions
break the $SU(2 N_f)$ symmetry into $SU(N_f) \otimes SU(N_f)$
at an order in $g_{V, A}$, while the higher derivative and
color magnetic interactions break the helicity
and the flavor symmetries at order ${\cal O}(1/E)$.
These new symmetries are expected to relate scattering amplitudes
of various channels each other in terms of
a universal elastic form factor, quite analogous
to the heavy quark symmetries \cite {wise}
utilized in the study of heavy quark weak decays.
In fact, as will be discussed later, the diffractive scattering
of constituent quarks is most precisely defined in terms of
high-energy boosted mesons containing heavy quarks.
The Feynman propagator of the quark $Q_{n \pm}$ is
found:
$$
< \!\! Q_{n \pm} (x) Q_{n' \pm}^\dagger (y) \!\!>
= \delta_{n, n'} P \exp [ i \!\!\int\!\! dx \cdot n \, (G_n + V_n \pm i
A_n)] \cdot \theta (x_o - y_o) \delta (n \cdot x - n \cdot y)
\delta^{(2)} (\vec x_\perp - \vec y_\perp)
\eqno (4)
$$
and similarly for the $Q_{\bar n\pm}$ quarks.
These propagators reveal features of the high-energy EFT
following from the new symmetries:
(1) the pair creation is not present,
i.e., the quark and and antiquark numbers are conserved as long as
their fluctuation is not large compared to the center of mass energy,
and (2) the quark wave function is simply phase-rotated in color and
flavor space by the external gluon and
the Goldstone boson fields along the scattering eikonal path.

For definiteness, consider the spin-polarized, $2 \rightarrow 2$
scatterings.
Denoting quantum numbers by $\alpha \beta \rightarrow \alpha' \beta'$,
the scattering matrix element is defined as:
$$
{\cal M}_{\rm el} (n \alpha, \bar n \beta; n' \alpha', \bar n' \beta')
\equiv {1 \over Z_n Z_{\bar n} }
<\!\! Q_{n' \alpha'} Q_{\bar n' \beta'} Q_{n \alpha}^\dagger
Q_{\bar n \beta}^\dagger \!\!>;
\hskip1cm Z_{n, \bar n} = <\!\! Q_{n \bar n} Q_{n, \bar n}  \!\!>.
\eqno (5)
$$
Of particular interest is the elastic, spin-polarized scattering
processes.
To project out the color and flavor octet channels while keeping
elastic singlet channel, we take a trace over the
color and flavor indices in Eq.(5) \cite {nachtman}. This yields
$$
{\cal M}_{\rm el} (\vec b) = \int d^2 \vec q e^{ i \vec q \cdot \vec b}
J_{\rm el}(q^2) ; \hskip1cm J_{\rm el} (q^2) = {1 \over Z_n Z_{\bar n}}
\int d^2 \vec b e^{-i
\vec q \cdot \vec b} < \!\! W_n (\vec b) W_{\bar n} (\vec 0) \!\!>
\eqno (6)
$$
in terms of the light-like Wilson-loops of helicity $\pm$:
$$
W_n^\pm (\vec x_\perp) \equiv {\rm Tr}_{\rm color\choose flavor}
P\exp[i\!\!\int\!\! G_{\bar n}
+ V_{\bar n} \pm A_{\bar n}] \equiv
{\cal V}_c (\vec x_\perp) \cdot {\cal V}^\pm_f (\vec x_\perp);
\eqno (7)
$$
$$
{\cal V}_c (\vec x_\perp)
\equiv {\rm Tr}_{\rm color} P \exp (i \!\! \int \!\! B_n);
\hskip1cm {\cal V}^\pm_f (\vec x_\perp)
\equiv {\rm Tr}_{\rm flavor} P \exp (i\!\!\int
\!\! V_n \pm A_n).
$$
The last equality in Eq.(7) expresses manifest {\elevenit factorization}
of the conventional `soft' and `hard' contributions to the elastic
scattering amplitudes within the present EFT.
However, it should be bear in mind that their relative strengths
vary continously as the momentum transfer $t$ is varied.
Therefore, the factorization Eq.(7) may not necessarily imply that
the two are distinct entity like the soft and the hard pomerons.
On the other hand, as the helicity dependence comes entirely
from the Goldstone boson exchanges, the factorization property might
be used to probe the relative soft and hard components of pomerons
by measuring experimentally helicity-even and -odd combinations of
the quark-quark diffractive scatterings.

Incidentally, the light-like Wilson loops are ill-defined due to
severe infared divergences once
the gluon and Goldstone boson dynamics are taken into account.
More precise derivation of the EFT getting around the problem is
in terms of the high-energy scattering between mesons consisting
a heavy quark \cite{bjorken} of mass $M_{\rm HQ}$.
Specifically, consider heavy meson - heavy
meson diffractive scattering. The heavy quark merely serves as an
eikonal trajectory for the brown mug of the light quarks.
As the mesons are boosted such that $\sqrt t \sim \Lambda_{\rm qcd}
< \!\! < M_Q <\!\!< \sqrt s$, the normalized
velocity 4-vector $n^\mu, \bar n^\mu$
of the brown mug, hence, the constituent quark becomes almost
light-like so that $n^2 = \bar n^2 \rightarrow 4 M_Q^2 / s$.
As the heavy quark's spin is irrelevant, the meson scattering then
{\elevenit defines} the constituent quarks' scattering
in Eq.(5) without any infrared divergences. Furthermore, this definition
has an obvious advantage of calculating the important {\elevenit s }
dependence of scattering amplitudes from the slight
timelike-ness $M_Q^2 / s$ of the scattering eikonal path.

The aforementioned Lorentz contraction of the longitudinal
gluon fields simplifies the gauge theory further. The contraction
may be understood as a result of anisotropic scaling \cite {verlinde}
to $ x^\mu \equiv (x_{||}, \vec b) \rightarrow (\lambda x_{||}, \vec b)
$ as $\lambda = { 2 M_Q / \sqrt s} \rightarrow 0$ while the magnitude
of {\elevenit s} is fixed.
In this limit, the longitudinally boosted gauge theory may be
expanded in powers of  $1/E$:
$$
L_{YM} \rightarrow {1 \over 4 \lambda^2 g^2} \sum_{||}tr G_{\mu \nu}^2
+ {1 \over 2 g^2} \sum_{|\perp}tr G_{\mu i }^2 + {\lambda^2  \over 4 g^2}
tr \sum_{\perp \perp} G_{ij}^2.
\eqno (8)
$$
Subscripts in the sum denote the plaquette orientations.
The Lagrangian implies that the gluons are strongly coupled along
the transverse, impact parameter two-dimensions. In the leading order
of $\lambda \propto \sqrt s$, the action is found reduced
to a $SU_c(3)$ chiral field theory on transverse two-dimensions
\cite {verlinde}:
$$
S_{VV} = \!\! \int \!\! d^4x
\, {1 \over 4 g^2} tr (
\partial_\mu ( U^\dagger D_i U))^2
= \int \!\! d^2 \vec x \, {1 \over 4 g^2} tr
(\nabla_i U^{\dagger A} K_{AB}
\nabla_i U^B)
\eqno (9)
$$
in which $A_\mu = U \partial_\mu U^\dagger$ using the constraint
$G_{\mu \nu} = 0$, and the kernel $K_{AB} = {+1 -1 \choose -1
+1}$.
It is then straightforward to evaluate the Wilson loop 2-point correlation
functions  in terms of the asymptotic global modes of the gauge
potential: $< \!\! W_n (G) W_{\bar n} (G) \!\!> = < tr U_{+\infty}(n)
U^\dagger_{-\infty}(n) tr U_{+\infty} (\bar n) U^\dagger_{-\infty}(\bar n)
>$. This yields the well-known eikonal result of the high-energy
quark scattering.
However, is is evident that the leading order effective theory Eq.(9)
cannot reproduce the
nonperturbative features of the pomerons.
The transverse fluctuation has to be taken into account properly,
whose leading logarithms was summed up in the hard pomeron case
\cite{lipatov2}.
As the transverse gluon interactions are strongly coupled, after
latticizing the transverse space, I propose
to replace Eq.(8) by a (gauged)
chiral field theory:
$$
L_{\rm YM} = {1 \over 4 \lambda^2 g^2} \sum_{||} tr G_{\mu \nu}^2
+ {1 \over 2} tr
\sum_{|\perp} |D_\mu M_i|^2 + {\lambda^2 g^2 \over 4} \sum_{\perp\perp}
tr (M_i M_j M_i^\dagger M_j^\dagger) + (h.c.) + \cdots
\eqno (10)
$$
In the singular limit $\lambda = 0$, the first term yields strictly
local color neutrality.
Alternatively, this may be viewed as an effective infrared action
of the anisotropic renormalization group transformation of the
bare Wilson's lattice theory in such a way that the longitudinal
dynamics has reached the weak coupling, continuum quasi-fixed point.
As both the kinematics and the boundary conditions are translationally
invariant along the scattering directions, the leading order theory
is effectively a two-dimensional QCD coupled to two adjoint scalar
matter field coming from the two transverse lattice link variables.
Fair amount of numerical studies by Bardeen et.al. \cite {transvqcd}
and, more recently, by Klebanov \cite {klebanov} indicated a
`Regge trajectory'-like spectrum (under suitable restrictions
to the Hilbert space) of this theory.
With the action Eq.(10), the light-cone Wilson loop
correlation functions can be evaluated through a mixed strong
and weak coupling expansions \cite {rey}
and the `Regge trajectory'-like spectrums are expected to
show up as poles in the t-channel.

Similarly, longitudinal gradients of the Goldstone
boson field are Lorentz contracted. One may integrate out the
eikonal fermions to obtain EFT of Goldstone bosons. Alternatively,
performing the similar rescaling as in the gluon dynamics, the
transverse EFT may be achieved. These two are related
each other roughly by a `gauge transformation'. The four-derivative terms
in Eq.(1) are then the leading order terms. It is seen that $V_n$
and $A_n$ approach a pure gauge at longitudinal infinities.
This implies that Goldstone boson field approaches to a fixed isospin
orientation, $\nabla_i \Sigma = 0$.
Utilizing this fact, the resulting Goldstone boson EFT is found to be
a $SU_f(3)$ chiral field theory on transverse two-dimensions:
$$
L_{\rm GB} \rightarrow
{\rm Tr} (\nabla_i \Sigma^{\dagger A} K_{AB} \nabla_i \Sigma^B) + \cdots
\eqno (11)
$$
The kernel $K_{AB}$ is the same as in the gluon EFT.
There is also an overall numerical factor depending on the
chiral Lagrangian coefficients $L_1,
\cdots$ in Eq.(1).
It is also possible to keep the subleading effects to Eq.(11).
and found an effective theory analogous to Eq.(10) is derived for
the Goldstone bosons as well.

In conclusion, an effective field theory of high-energy diffractive
scattering is derived utilizing the dynamics of constituent quarks
with gluons and Goldstone bosons. To the leading order, it is shown
that the gluon and Goldstone boson dynamics is captured by two
separate chiral field theories of $SU(3)_c$ and $SU(3)_f$ degrees
of freedom. Spin polarized elastic scattering amplitudes are then
obtained as quantum average of products of two-point, light-like Wilson
loop correlation functions using the chiral field theories.
The gluon and the Goldstone boson exchanges are then distinguishable
by looking at various combinations of the spin-polarized quark-quark
(or heavy meson - heavy meson) diffractive scatterings.

I thank L.N. Lipatov, V.P. Nair, S. Nussinov, M. Shifman and
C.-I. Tan for many helpful discussions. This work was supported
in part by KOSEF-SRC Program, Ministry of Education Grant through
RIBS, and Daewoo Foundation.

\end{document}